\documentclass[12pt]{article}

\usepackage{epsfig}
\usepackage{latexsym}
\def\balpha{\mbox{\boldmath $\alpha$}}
\def\btau{\mbox{\boldmath $\tau$}}
\def\bnu{\mbox{\boldmath $\nu$}}
\def\bpsi{\mbox{\boldmath $\psi$}}
\def\half{\textstyle{\frac{1}{2}}}
\input{amssymb.sty}

\begin{document}

\begin{titlepage}

\baselineskip 24pt

\begin{center}

{\Large {\bf A Solution of the Strong CP Problem Transforming the 
theta-angle to the KM CP-violating Phase}}

\vspace{.5cm}

\baselineskip 14pt

{\large Jos\'e BORDES\footnote{work supported by
Spanish MICINN under contract FPA2008-02878}}\\
jose.m.bordes\,@\,uv.es\\
{\it Departament Fisica Teorica and IFIC, Centro Mixto CSIC,
\\Universidad de
Valencia\\
  calle Dr. Moliner 50, E-46100 Burjassot (Valencia), Spain}\\
\vspace*{.4cm}
{\large CHAN Hong-Mo}\\
hong-mo.chan\,@\,stfc.ac.uk\\
{\it Rutherford Appleton Laboratory,\\
  Chilton, Didcot, Oxon, OX11 0QX, United Kingdom}\\
\vspace{.4cm}
{\large TSOU Sheung Tsun}\\
tsou\,@\,maths.ox.ac.uk\\
{\it Mathematical Institute, University of Oxford,\\
  24-29 St. Giles', Oxford, OX1 3LB, United Kingdom}

\end{center}

\vspace{.3cm}

\begin{abstract}

It is shown that in the scheme with a rotating fermion mass 
matrix  (i.e.\ one with a scale-dependent orientation in 
generation space) 
suggested earlier for explaining fermion mixing and mass 
hierarchy, the theta-angle term in the QCD action of
topological origin can be eliminated by chiral
transformations, while giving still nonzero masses to all quarks.  
Instead, the effects of such transformations get transmitted by the
rotation to the CKM matrix as the KM phase giving, for $\theta$ of order
unity, a Jarlskog invariant typically of order $10^{-5}$ as
experimentally observed.  Strong and weak CP violations appear then as
just two facets of the same phenomenon.

\end{abstract}

\end{titlepage}

\clearpage

\baselineskip 14pt

A long-standing puzzle in particle theory is the so-called strong
CP problem.  Lorentz and gauge invariance allow in principle in 
the QCD action a term of the form:
\begin{equation}
{\cal L}_{\theta} = - \frac{\theta}{64 \pi^2} 
   \epsilon^{\mu \nu \rho \sigma} F_{\mu \nu} F_{\rho \sigma}
\label{Ltheta}
\end{equation}
of topological origin, where $\theta$ can take any arbitrary value.
Experiment, on the other hand, does not seem to admit such a term
which violates CP invariance and can lead to a nonvanishing electric 
dipole moment for the neutron of the order \cite{Weinberg}:
\begin{equation}
d_n \sim |\theta| e m_\pi^2/m_N^3 \sim 10^{-16} |\theta| e \  cm.
\label{edm}
\end{equation}
The experimental limit for this has by now been pushed to below
$2.7 \times 10^{-26} e \ cm$ \cite{edm}, which means that $|\theta|$ 
has to have a value below $3 \times 10^{-10}$.  It would thus seem 
that nature has a reason unknown to us either for suppressing this 
term to such a small value, or else for eliminating it altogether.

The favourite candidate among theoreticians for explaining away 
the theta-angle term is the axion theory \cite{PecceiQuinn,
Weinberg1,Wilczek}, which axions, however, have been diligently 
searched for in experiment since first suggested, i.e. for over 
forty years, yet not been found.

Now, it has long been known that the theta-angle term can be 
eliminated if there are quarks of zero mass.  Effecting a chiral 
transformaton on a quark field, thus:
\begin{equation}
\psi \longrightarrow \exp(i \alpha \gamma_5) \psi
\label{chiraltrans}
\end{equation}
will yield a term in the Feynman integral of the same form as the
theta-angle term.  Hence, if we make a chiral transformation
(\ref{chiraltrans}) on each quark flavour, we shall end up with
a theta-angle term modified to:
\begin{equation}
\theta \longrightarrow \theta + 2 \sum_{F} \alpha_F,
\label{modtheta}
\end{equation}
which can be made to vanish by a judicious choice of $\alpha_F$.
However, a chiral transformation on a massive fermion field will 
in general make its mass parameter complex:
\begin{eqnarray}
m \bar{\psi} \psi 
   & = & m \bar{\psi} \half (1 + \gamma_5) \psi
   + m \bar{\psi} \half (1 - \gamma_5) \psi \nonumber\\ 
   & \rightarrow &
   m \exp(2i \alpha) \bar{\psi} \half (1 + \gamma_5) \psi
 + m \exp(- 2i \alpha) \bar{\psi} \half (1 - \gamma_5) \psi,
\label{modmass}
\end{eqnarray}
and lead again to CP-violations.  Only when the quark has a zero 
mass can such a conclusion be avoided.  Unfortunately, none of
the quarks known can be assigned a zero mass in experiment, and 
so the problem remains.

A parallel problem actually also exists in the weak sector although 
it is not usually considered as such.  General invariance principles 
there imply that the weak current is of the form:
\begin{equation}
J^\mu = \bar{\left( \begin{array}{c} u \\ c \\ t \end{array} 
   \right)} \gamma^\mu (1 - \gamma_5) V \left( \begin{array}{c}
   d \\ s \\ b \end{array} \right),
\label{wkcurrent}
\end{equation}
where $V$, the CKM mixing matrix relating the $U$-type to $D$-type
quarks, depends on 4 parameters, one of which is the Kobayashi-Maskawa
phase \cite{KM} which violates CP.  In contrast to the theta-angle 
term in the strong sector, the effects of the KM phase here have been 
detected in experiment, but again are suppressed by nature for reasons 
unknown, though not to the same drastic extent.  The Jarlskog invariant
\cite{Jarlskog} $J$ which is a convenient measure of the CP-violating
effects here is found to have a value of only \cite{Databook}:
\begin{equation}
J \sim 3 \times 10^{-5}.
\label{Jarls}
\end{equation} 
One could well call this in parallel the ``weak CP problem'', only here, 
the problem being bound up with the mixing angles, i.e.\ the 3 other
``real'' parameters in the matrix $V$, the small values of some of which 
are equally unexplained in the usual formulation of the Standard Model, 
one is not so inclined to label it as such.  Nevertheless, this ``weak 
CP problem'' is quite as puzzling as the strong, and has remained with 
us almost as long. 

In view of these problems, we were interested first to note recently 
\cite{strongcp} that in the framework of a rotating mass matrix (i.e.\
one which has an orientation in generation space which is 
scale-dependent) we 
have constructed earlier \cite{phenodsm} to explain fermion mixing 
and mass hierarchy, there is a possible solution to the strong CP 
problem by means of chiral transformations which, nevertheless, 
because of the rotation, preserve the hermiticity of the mass matrix 
and give nonzero masses to all quarks.  Then subsequently, following 
through these earlier considerations down to the mixing matrix level, 
we have since found, as we shall show below, that the elimination of 
the theta-angle term in the strong sector automatically leads to a 
mixing matrix with a Kobayashi-Maskawa phase, the CP-violating effects 
of which, being themselves contingent on the low speed of the rotation, 
will naturally yield, for $\theta$ of order unity, small values for the
Jarlskog invariant, typically of the order $10^{-5}$ as observed in 
experiment.  In other words, the rotating mass matrix framework seems 
to offer a simultaneous solution to both the strong and ``weak'' CP 
problems, while linking the two intriguingly together - and these in 
addition to its original offer of an explanation for fermion mixing 
and mass hierarchy.   

That a rotating fermion mass matrix can give rise to both mixing 
and mass hierarchy is in itself a very simple idea which can easily 
be seen as follows.  One starts with a fermion mass matrix of the 
usual form:
\begin{equation}
m \half(1 + \gamma_5) + m^{\dagger} \half(1 - \gamma_5),
\label{m}
\end{equation}
which, following Weinberg \cite{Weinberg2}, one can always rewrite
by a relabelling of the singlet right-handed fields, with no change
in physics, in a hermitian form independent of $\gamma_5$, a form
we shall henceforth adopt.  Suppose now this matrix is factorizable,
meaning that it is of the form:
\begin{equation}
m = m_T \balpha \balpha^{\dagger},
\label{mfact}
\end {equation}
where \balpha\ is a normalized
global (i.e.\ $x$-independent) vector in
generation space.  We can even suppose that \balpha\  is 
universal and that only the numerical coefficient $m_T$ depends on 
the fermion type.  Obviously, such a mass matrix has only one massive 
state represented by the vector \balpha, and zero mixing, i.e.\ 
only the identity matrix as the mixing matrix.  This is not unattractive 
as a starting point for quarks, as has occurred already a long time ago 
to several authors \cite{Fritsch,Harari}, but is clearly insufficiently 
realistic in detail.

However, if one now says that the vector \balpha\ rotates with
changing scale $\mu$, as proposed, then the situation becomes very
interesting.  All quantities now depend on the scale $\mu$ and one 
has to specify at which scale the mass or state vector of each
particle is to be measured.  Suppose one follows the usual convention
and defines the mass of each particle as that measured at the scale 
equal to its mass, we find then that mixing and mass hierarchy would
immediately result.  

\begin{figure} [ht]
\centering
\input{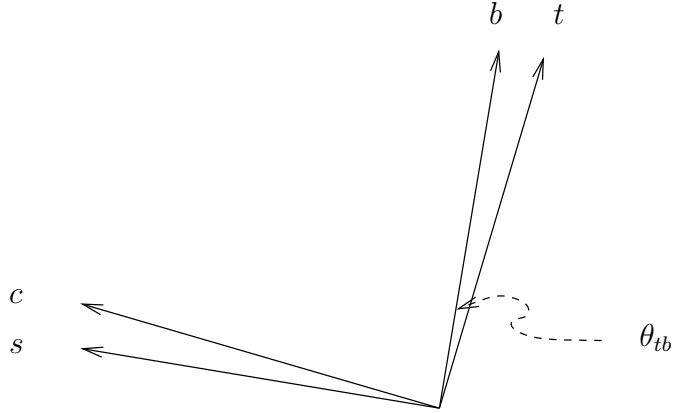}
\caption{Mixing between up and down fermions from a rotating mass matrix.}  
\label{UDmix}
\end{figure}

To see how this comes about, it is sufficient for illustration to 
consider a situation where one takes account only of the two heaviest 
states in each fermion type.  By (\ref{mfact}) then, taking for the 
moment \balpha\ to be real and $m_T$ to be $\mu$-independent for 
simplicity, we would have $m_t = m_U$ as the mass of $t$ and the 
eigenvector $\balpha(\mu = m_t)$ as its state vector ${\bf v}_t$.  
Similarly, we have $m_b = m_D$ as the mass and $\balpha (\mu = m_b)$ 
as the state vector ${\bf v}_b$ of $b$.  Next, the state vector 
${\bf v}_c$ of $c$ must be orthogonal to ${\bf v}_t$, $c$ being by 
definition an independent quantum state to $t$.  Similarly, the state 
vector ${\bf v}_s$ of $s$ is orthogonal to ${\bf v}_b$.  
So we have the 
situation as illustrated in Figure \ref{UDmix}, where the vectors 
${\bf v}_t$ and ${\bf v}_b$ are not aligned, being the vector 
$\balpha (\mu)$ taken at two different values of its argument $\mu$, 
and \balpha\  by assumption rotates.  This gives then the following 
CKM mixing (sub)matrix, in the situation considered with only the two 
heaviest states:
\begin{equation}
\left( \begin{array}{cc} V_{cs} & V_{cb} \\ V_{ts} & V_{tb} \end{array}
   \right) = \left( \begin{array}{cc} \langle {\bf v}_c|{\bf v}_s \rangle 
                             & \langle {\bf v}_c|{\bf v}_b \rangle \\
                               \langle {\bf v}_t|{\bf v}_s \rangle 
                             & \langle {\bf v}_t|{\bf v}_b \rangle
             \end{array} \right )
           = \left( \begin{array}{cc} \cos \theta_{tb} & \sin \theta_{tb} \\  
                -\sin \theta_{tb} & \cos \theta_{tb} \end{array} \right),    
\label{UDmixing}
\end{equation}
which is no longer the identity, hence mixing.

\begin{figure} [ht]
\centering
\includegraphics{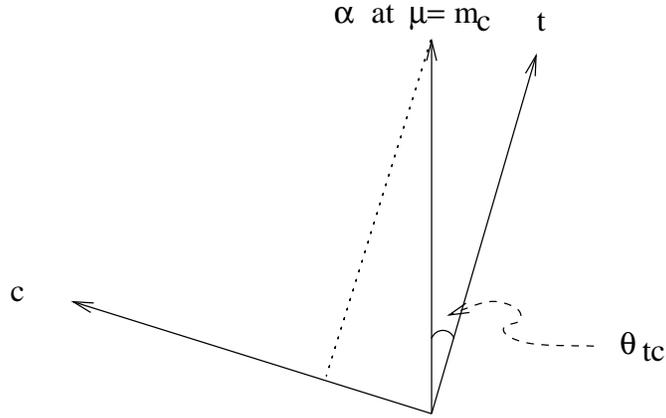}
\caption{Masses for lower generation fermions from a rotating mass 
matrix.}  
\label{alphaU}
\end{figure}

Next, what about hierarchical masses?  From (\ref{mfact}), it follows 
that ${\bf v}_c$ must have zero eigenvalue at $\mu = m_t$.  But this 
value is not to be taken as the mass of $c$ which we agreed has to be 
measured at $\mu = m_c$.  In other words, $m_c$ is instead to be taken 
as the solution to the equation:
\begin{equation}
\mu = \langle {\bf v}_c|m(\mu)|{\bf v}_c \rangle 
    = m_U |\langle {\bf v}_c| \balpha (\mu) \rangle|^2.
\label{solvmc}
\end{equation}
A nonzero solution exists since \balpha\  by assumption rotates
so that at $\mu \neq m_t$, it would have rotated to some direction
different from ${\bf v}_t$, as illustrated in Figure \ref{alphaU}, 
and acquired a component $\sin \theta _{tc}$ in the direction 
of ${\bf v}_c$ giving thus:
\begin{equation}
m_c = m_t \sin^2 \theta_{tc},
\label{mc}
\end{equation}
which is nonzero, but will be small if the rotation is not too fast,
hence mass hierarchy.

Clearly, the same arguments can be applied to the full 3-generation
case, to down quarks, and to leptons as well, although with the last  
we shall not here be directly concerned.  The question, however,
is whether the mixing matrices and hierarchical masses so obtained
are anything like those experimentally observed, and whether viable
theoretical models can be constructed to reproduce the required
rotation.  This question we have studied for some time.  The first 
point is adressed in \cite{cevidsm,phenodsm}, for example, and the 
second in \cite{genmixdsm,prepsm}, to which the interested reader 
is referred for a summary, and to references therein for details, 
of the results so far obtained.

As regards the CP problem of present interest, the above scenario
has one quite striking feature in that, despite the conclusion 
that all fermions have nonzero, though hierarchical, masses the
mass matrix that appears in the action remains factorizable (rank 1)
for all scales.  This comes about basically from unitarity which 
means that the physically measured masses of the two lower generation
fermion states are not the eigenvalues of the mass matrix appearing 
in the action but those of some truncations of that mass matrix 
\cite{strongcp}.  Thus, as far as the mass matrix appearing in the 
action is concerned, there are still directions in generation space, 
namely on the plane orthogonal to \balpha, in which the 
fermion field has zero eigenvalues.  And in those directions, chiral 
transformation can still be performed on the fermion field without 
affecting the hermiticity of the mass matrix, as emphasized after 
(\ref{modmass}) above.  In other words, for a mass matrix of rank 1,
the relabelling of right-handed fields required to make it appear 
hermitian as in (\ref{mfact}) is non-unique, and leaves sufficient
freedom to allow some chiral transformations on the fermion fields.
Hence, by a judicious choice of these, any theta-angle term in the 
action can be eliminated.  This point was noted in \cite{strongcp} 
already, but that was as far as we had got there.

Our intention now is to follow through the argument with rotation 
and see what other information this will give us.  In doing so, we 
shall retain the simplifying assumption of a real \balpha\ 
made above for ease of illustration.  We keep the assumption now, 
however, deliberately, on purpose to demonstrate that even for 
\balpha\  real, a CP-violating Kobayashi-Maskawa phase will 
authomatically appear in the CKM matrix when the theta-angle is 
eliminated.  Only at the end shall we return to discuss what 
happens when \balpha\ becomes complex.  

At each scale $\mu$, there are then two independent directions with 
zero eigenvalues to the mass matrix $m$, namely the two directions on 
the plane orthogonal to $\balpha(\mu)$.  It would thus seem that 
chiral transformations can be effected on any two of these directions
without affecting the hermiticity of the mass matrix so as to eliminate 
the theta-angle term from the action at $\mu$.  On closer examination, 
however, this does not appear to be the case when rotation is involved.  
The action at $\mu$ prescribes not only the vector \balpha\ at 
$\mu$, but also, via renormalization, the vector \balpha\ at 
$\mu + d \mu$, and it is this bit of information that on iteration 
gives us the rotation trajectory for \balpha.  At both $\mu$
and $\mu + d \mu$, however, the mass matrix is supposed to be of
the form (\ref{mfact}), and therefore hermitian.  Hence, in performing 
a chiral transformation to eliminate the theta-angle term at $\mu$, 
we would 
want not only to keep hermitian the mass matrix at $\mu$, but 
also the mass matrix at $\mu + d \mu$.  We notice, however, that 
although the mass matrix at $\mu$, namely $m(\mu) = m_T \balpha(\mu) 
\balpha(\mu)^{\dagger}$ has zero eigenvalue in the direction 
$\dot{\balpha}(\mu)$, this being orthogonal to $\balpha(\mu)$, 
the mass matrix at the neighbouring point $\mu + d \mu$, namely:
\begin{equation}
m(\mu + d \mu) = m_T [\balpha(\mu) + \dot{\balpha}(\mu) 
   d\mu] [\balpha(\mu) + \dot{\balpha}(\mu) d\mu]^{\dagger}
\label{mmu+dmu}
\end{equation}
has a nonzero value in the direction $\dot{\balpha}(\mu)$.  Thus,
a chiral transformation in the direction $\dot{\balpha}(\mu)$ will
not leave the mass matrix $m(\mu + d \mu)$ hermitian.  Only a chiral 
transformation in the direction orthogonal to both $\balpha(\mu)$ 
and $\dot{\balpha}(\mu)$ will leave both the matrices $m(\mu)$ and 
$ m(\mu + d \mu)$ hermitian.

Let us set up then at each point $\mu$ of the rotation trajectory a
(Darboux) triad, namely the radial vector $\balpha(\mu)$, the
tangent vector $\btau(\mu) = \dot{\balpha}(\mu)$ and the normal
vector $\bnu(\mu)$ orthogonal to both.  The conclusion of the
preceding paragraph is that the chiral tansformation to eliminate
the theta-angle term should be effected in the direction $\bnu (\mu)$ 
at every $\mu$.  To be explicit, let us choose a reference frame in
3-D generation space such that at $\mu = \infty$, we have:
\begin{equation}
\balpha(\infty) = \balpha_0 = (1, 0, 0)^\dagger; \ \ 
   \btau(\infty) = \btau_0 = (0, 1, 0)^\dagger; \ \ 
   \bnu(\infty) = \bnu_0 = (0, 0, 1)^\dagger,
\label{Barbouxinf}
\end{equation}
and define a rotation $A(\mu)$ such that:
\begin{equation}
\balpha(\mu) = A(\mu) \balpha_0; \ \ 
   \btau(\mu) = A(\mu) \btau_0; \ \ 
   \bnu(\mu) = A(\mu) \bnu_0.
\label{Amu}
\end{equation}
The chiral transformation needed 
at scale $\mu$ can then be represented as:
\begin{equation}
P(\mu) = A(\mu) P_0 A^{-1}(\mu), \ \ P_0 = \left( \begin{array}{ccc}
 1 & 0 & 0 \\ 0 & 1 & 0 \\ 0 & 0 & e^{- i \theta \gamma_5/2}
 \end{array} \right).
\label{Pmu}
\end{equation}
Effecting such a chiral transformation on the 3-component
quark field \bpsi\ (for 3 generations) at every
$\mu$ removes the theta-angle term entirely and leave the
action CP-invariant and the mass matrix hermitian at every $\mu$.

Notice, however, that although by choosing to perform the chiral
transformation in the normal direction $\bnu(\mu)$ at scale 
$\mu$, we have ensured that the mass matrix is hermitian not only
at $\mu$ but also at a neighbouring $\mu + d\mu$, it will generally
make $m$ non-hermitian at $\mu'$ for $\mu - \mu'$ finite, since 
\bnu\ would have rotated at $\mu'$ to a different direction 
giving then the chiral transformation $P(\mu)$ there a massive
component and hence a non-hermitian mass matrix.  Explicitly, under a
chiral transformation $P(\mu)$, the mass term at scale $\mu'$, in
close parallel to (\ref{modmass}), would transform as:
\begin{eqnarray}
\lefteqn{
m_T \bar{\bpsi} \balpha(\mu') \balpha(\mu')^\dagger
   \bpsi } \nonumber\\
&& \!\!\!\rightarrow m_T \bar{\bpsi} A(\mu) P_0 A^{-1}(\mu)
   \balpha(\mu') \balpha(\mu')^\dagger \half(1+\gamma_5)
   A(\mu) P_0 A^{-1}(\mu) \bpsi \nonumber\\
&&\!\!\!{}+  m_T \bar{\bpsi} A(\mu) P_0 A^{-1}(\mu)
   \balpha(\mu') \balpha(\mu')^\dagger \half(1-\gamma_5)
   A(\mu) P_0 A^{-1}(\mu) \bpsi.
\label{modMass}
\end{eqnarray}
If $\mu' = \mu$, then $A^{-1} \balpha = (1, 0, 0)^\dagger$, in 
which case $P_0$ operating on it will leave it invariant.  But for
$\mu' \neq \mu$, this no longer applies, and $P_0$ will give in
general a phase which is different in the 2 terms after the
transformation 
and make the mass matrix non-hermitian.  Notice that although
this mass matrix is still, of course, of the general form (\ref{m}) 
we started with, we are not allowed now to relabel the right-handed 
fields to recast it in a hermitian form as we did before in deriving 
(\ref{mfact}), for this would change the relative phases between the
right- and left-handed fields, which is equivalent to a chiral
transformation and can therefore resuscitate the theta-angle term 
that we have taken such care earlier to eliminate.  To recover a
hermitian mass matrix and yet eliminate the theta-angle term, we
shall need to first undo the chiral transformation performed above
at $\mu$ and then perform the chiral transformation again at $\mu'$ 
instead.  In other words, we need to apply to the quark field the 
operator $P(\mu') P^{-1}(\mu)$ to obtain the desired result.

Indeed, the chiral transformation $P(\mu)$ guarantees only that the 
mass matrix is hermitian at the two neighbouring points $\mu$ and
$\mu + d \mu$.  If we wish to iterate the procedure so as to 
eliminate the theta-angle term at $\mu + d \mu$ while ensuring the 
hermiticity of the mass matrix at the next neighbouring point too,
then we have first to undo the chiral transformation performed at
$\mu$ before, and then effect again the chiral transformation at
$\mu + d \mu$, namely to apply the operator
\begin{equation}
P(\mu + d \mu) P^{-1}(\mu)  
\label{paratrans}
\end{equation}
to the quark fields, in order to obtain the desired result.  This 
operator (\ref{paratrans}) acts thus as a sort of parallel transport, 
detailing effectively what is meant by the same or parallel (chiral) 
phases at two neighbouring points along the trajectory, and hence, 
by iteration, at any two points a finite distance apart as detailed
in the paragraph above.      

To see how these chiral transformations will affect the conclusions
above on quark masses and mixing, let us start with just one species 
of quarks, say, the U-type quarks, i.e.\ $t, c, u$.  The state vector 
of $t$, i.e.\ ${\bf v}_t$, or just ${\bf t}$ for short, is defined as 
$\balpha(\mu = m_t)$ and the state vectors ${\bf c}, {\bf u}$ are 
to be orthogonal to it and are themselves mutually orthogonal.  It 
follows therefore that the dyad ${\bf c},{\bf u}$ is related to the
dyad ${\btau}(\mu = m_t) = \btau_U, \bnu(\mu = m_t) = \bnu_U$ 
just by an orthogonal transformation, thus:
\begin{equation}
\btau_U = \Omega_U {\bf c}; \ \ \bnu_U = \Omega_U {\bf u},
\label{taunutu}
\end{equation}
with
\begin{equation}
\Omega_U = \left( \begin{array}{ccc} 1 & 0 & 0 \\ 
                  0 & \cos \omega_U & - \sin \omega_U \\
                  0 & \sin \omega_U & \cos \omega_U \end{array} \right),
\label{Omega}
\end{equation} 
$\omega_U$ being just the angle between ${\bf c}$ and $\btau_U$.
This angle is small but nonzero, since
${\bf c} = {\bf v}_c$ is the vector which is orthogonal to
$\balpha(\mu = m_t)$ and lies on the plane containing both the 
vectors $\balpha(\mu = m_t)$ and $\balpha(\mu = m_c)$ \cite{phenodsm}, 
while $\btau_U$ is the tangent to the trajectory at $\mu = m_t$; 
it is thus a measure of how much $\balpha(\mu)$ has rotated from
$\mu = m_t$ to $\mu = m_c$.  

Suppose we wish again to evaluate the mass of the $c$ quark in the
rotation scenario as we did before but incorporating now the above
procedure for eliminating the theta-angle term.  We agreed that this
has to be done at the scale $\mu=m_c$ so that all scale-dependent
quantities involved should be evaluated at this scale also.  Thus,
with the $c$ state vector ${\bf c}$ defined originally at $\mu=m_t$,
the $c$ quark field is given there as $\psi_c (\mu=m_t)=
{\bf c}^\dagger \cdot P(m_t) \bpsi$, but it will now have to be parallelly
transported by (\ref{paratrans}) to $\mu=m_c$, giving instead $\psi_c
(\mu=m_c) = {\bf c}^\dagger \cdot P(m_c) \bpsi$.  The mass term also,
according to (\ref{modMass}) above, will now appear as:
\begin{equation}
m_T \bar{\bpsi} P(m_c) \balpha (m_c) \balpha^\dagger (m_c) P(m_c)
\bpsi,
\label{modMc}
\end{equation}
where the operators $P(m_c)$ can in fact be omitted since the vector
$\balpha (m_c)$ is invariant under $P(m_c)$.  What interests us here
as far as the $c$ mass is concerned, according to the analysis in, for
example, \cite{strongcp}, is the diagonal contribution from the $c$
quark, namely:
\begin{eqnarray}
\lefteqn{
m_T  \bar{\bpsi} P(m_c) {\bf c} {\bf c}^\dagger \balpha (m_c)
\balpha^\dagger (m_c) {\bf c} {\bf c}^\dagger P(m_c) \bpsi} \nonumber \\
 &&= m_T | {\bf c}^\dagger \cdot \balpha (m_c)|^2 \bar{\psi}_c (\mu=m_c)
\psi_c (\mu=m_c),
\label{ccontrm}
\end{eqnarray}
giving then the $c$ mass as:
\begin{equation}
m_c=m_T | {\bf c}^\dagger \cdot  \balpha (m_c)|^2,
\label{mcagain}
\end{equation}
i.e., exactly the same as before when no consideration was given to
the elimination of the theta-angle term, as in (\ref{mc}) above for
the simplified 2-generation version.

The same considerations will apply also to $m_u$ as for $m_c$.
One sees therefore 
that so long as there is only one type of quarks, one can always manage, 
with a rotating factorizable mass matrix, to eliminate any theta-angle 
term so as to maintain CP-conservation, while keeping the mass matrix 
hermitian, and having at the same time hierarchical but nonzero masses 
for all quarks. 

What happens, however, when there are both up-type and down-type 
quarks?  In that case, the two types can be coupled by the weak 
current via the CKM mixing matrix, and one has again to follow through 
the preceding arguments and trace out the consequence of eliminating 
the theta-angle term.  To do so, let us denote the state vectors of 
the U-type quarks defined above at $\mu = m_t$ together as:
\begin{equation}
V_U = ({\bf t}, {\bf c}, {\bf u}) = \left( \begin{array}{ccc}
   t_1 & c_1 & u_1 \\ t_2 & c_2 & u_2 \\ t_3 & c_3 & u_3
   \end{array} \right),
\label{VU}  
\end{equation}
and similarly the state vectors of the D-type quarks defined at
$\mu = m_b$ as:
\begin{equation}
V_D = ({\bf b}, {\bf s}, {\bf d}) = \left( \begin{array}{ccc}
   b_1 & s_1 & d_1 \\ b_2 & s_2 & d_2 \\ b_3 & s_3 & d_3
   \end{array} \right),
\label{VD}  
\end{equation}
where in the notation introduced above, we have from rotation:
\begin{eqnarray}
V_U & = & A_U \Omega_U; \ \  A_U = A(\mu = m_t); \nonumber\\
V_D & = & A_D \Omega_D; \ \  A_D = A(\mu = m_b).
\label{VUVD}
\end{eqnarray}
What interests us is the relative orientation of $V_U$ and $V_D$,
the matrix of inner products between the state vectors of the
U-type and D-type quarks being the CKM mixing matrix we seek.  In
order to compare the orientation of the state vectors of one type
to those of the other, the two types being defined as they are at
two different scales, we need first to ``parallelly transport'' the 
(chiral) phase of each to the same scale.  Chiral transformations 
should, of course, be performed in principle on the quark fields, 
which can be done directly to the weak current in (\ref{wkcurrent}),
but since in (\ref{wkcurrent}) only 
left-handed fields are involved, the factor $\exp (-i \theta \gamma_5)$ 
in all chiral transformations can be replaced just by the phase 
factor $\exp (i \theta)$ and can thus
be conveniently transferred from the 
quark fields to their state vectors for ease of presentation.  
Parallelly transporting then to a common scale, say $X$, for comparison,
we have:
\begin{eqnarray}
P_X P_U^{-1} V_U & = & P_X (A_U P_0^{-1} A_U^{-1})(A_U \Omega_U)
   = \tilde{V}_U; \\ \nonumber
P_X P_D^{-1} V_D & = & P_X (A_D P_0^{-1} A_D^{-1})(A_D \Omega_D)
   = \tilde{V}_D.
\label{VIVDtilde}
\end{eqnarray}
Hence we obtain the CKM matrix in this scenario as:
\begin{equation}
V_{CKM} = \tilde{V}_U^{-1} \tilde{V}_D
        = (\Omega_U^{-1} P_0 \Omega_U) V_U^{-1} V_D
          (\Omega_D^{-1} P_0^{-1} \Omega_D),
\label{VCKM}
\end{equation}
where the factors $P_X$ cancel, meaning that it does not matter at 
which common scale we choose to make the comparison, as expected.

Notice that had there been no theta-angle term to contend with, we
would have obtained for the CKM matrix just the factor $V_U^{-1} 
V_D$, which will be a real matrix with no CP-violating phase in 
it, having started with a real \balpha\ as we have done.  By 
insisting on the chiral transformations to eliminate the theta-angle 
term throughout, we have then injected some new phases into the 
mixing matrix elements, and hence the possibility of CP-violation,
which will be the case if the phases introduced by the said chiral
transformations cannot be removed by any changes in phase of the 
quark states.  Indeed, if we were to put $\omega_U = \omega_D= 0$,
with $\omega_D$ similarly defined as $\omega_U$  
in (\ref{Omega}),
we would have obtained a vanishing value for the
Jarlskog invariant and 
no CP-violation.  The reason is clear, since in that case the vector 
${\bf u}$ would coincide with the normal vector \bnu\ at 
$\mu = m_t$ , and $\bf d$ with \bnu\ at $\mu=m_d$,
on which the chiral transformations are performed, and 
the effect on the CKM matrix would be the same as that of changing 
the phases of the $u$ and $d$
fields, which are in any case arbitrary.  If we
were to calculate the Jarlskog invariant from (\ref{VCKM}) in this 
case, the phases would cancel and one obtains a zero value.  Since,
however, $\omega_U$ and $\omega_D$
are nonzero by virtue of the rotation as explained 
above, this cancellation has now no reason to occur and one has 
in general nonvanishing Jarlskog invariants and CP-violations as 
the result.

But will this yield Jarlskog invariants and CP-violating effects of
the order observed in experiment?  We recall that the strong CP
angle $\theta$ from which this effect is supposed to originate can
take in principle any arbitrary value and so should be taken,
without prejudice, as of
order unity, whereas the measured value of the Jarlskog invariant
is of order $3 \times 10^{-5}$ \cite{Databook}, so that a suppression 
by about 4 
orders of magnitude is required in the process of transmitting the
CP-violating effects from the strong sector to the weak sector via 
rotation.  This is possible, so long as the rotation is relatively 
slow as is envisaged.  To see whether it is indeed the case, one 
can evaluate the Jarlskog invariant for (\ref{VCKM}) with, for 
example, its $2 \times 2$ submatrix labelled by the 2 heaviest 
states $t, c$ and $b, s$.  One obtains then an explicit expression 
for $J$ in terms of $\theta$, $\omega_U$, $\omega_D$ and elements of 
the matrix $V_U^{-1} V_D$.  The angles $\omega_U$ and $\omega_D$, 
one has already noted to be of order $\epsilon$, the angle rotated 
by the vector \balpha\ from the scale of the heaviest to that
of the second generation.  Further, from an earlier analysis of the
rotation picture \cite{features}, one has learned that the CKM 
matrix elements $V_{ts}, V_{cb}, V_{cd}, V_{us}$ proportinal to the
curvatures of the rotation trajectory are all of order $\epsilon$, 
while the corner elements $V_{td}, V_{ub}$ proportional to its 
torsion are of order $\epsilon^2$.  This is not to say, of course,
that all four elements of order $\epsilon$ need be of the the same
size, for the two curvatures, normal and geodesic, can have quite 
different values, as seems to be the case for a trajectory fitted
to experiment.  But, for the order-of-magnitude estimate generic
to the rotation scheme that we aim for at the moment, we shall 
deliberately ignore such details specific to a particular trajectory.  
Substituting then the above estimates into the formula for $J$, 
one finds that $J$ is of order $\epsilon^4$ and proportional to 
$\sin (\theta/2)$.  An estimate for the value of $\epsilon$ can be 
obtained from the rotation formula (\ref{mc}) for the mass ratio 
of the second generation to the heaviest, leading to $\epsilon 
\sim 0.08$ from $m_c/m_t$, and $\epsilon \sim 0.14$ from $m_s/m_b$, 
\cite{Databook}.  This then gives an order-of-magnitude estimate 
for the Jarlskog invariant as:
\begin{equation}
J \sim \sin (\theta/2) \times 10^{-4},
\label{Jorder}
\end{equation}
which is quite consistent with the experimentally measured value 
(\ref{Jarls}) for a strong CP angle $\theta$ of order unity

One knows of course, whether in the rotation framework or otherwise, 
that once given the small values observed in experiment for mixing 
angles involving the two heaviest states $t$ and $b$, it will follow 
already that the CP-violating effects of the KM phase will be small, 
since it is known that for two generations there is no CP-violation, 
which can thus arise only through mixing with $t$ and $b$.  A priori,
however, one can give no actual estimate for the size of the effect, 
not knowing how or where the KM phase originates.  The difference 
with the scheme here is that, first, having traced the origin of 
the KM phase via rotation back to the strong sector, one can give 
now an actual estimate for $J$, and second, since the rotation 
relates also the mixing angles of fermions to their hierarchical 
masses, as explained in (\ref{UDmixing}) and (\ref{mc}) above, the 
estimate can be derived with only mass ratios as inputs, with no 
empirical knowledge of the mixing angles being required at all.
  
If one is willing further to supply the experimentally measured 
values of the mixing angles as inputs, then 
one would obtain in the rotation scheme
a more accurate estimate of the Jarlskog invariant.
Thus, if we take for $V_U^{-1} V_D$ in (\ref{VCKM}) the matrix of 
absolute values of CKM elements given in \cite{Databook}, one is 
left only with $\omega_U$ and $\omega_D$ as unknowns.  Both these 
angles, however, are seen to be related to the geodesic curvature of 
the rotation trajectory \cite{features}, which is itself related 
to the Cabibbo angle.  Hence, reasoning along these lines, and 
inputting the measured value $\sim 0.22$ of the Cabibbo angle, 
one obtains the following rough
estimates: $\omega_U \sim 0.16$ and 
$\omega_D \sim 0.28$, which when substituted into (\ref{VCKM}) 
give:
\begin{equation}
J \sim \sin (\theta/2) \times 7.7 \times 10^{-5}. 
\label{Jestim}
\end{equation}
This will coincide with the experimentally measured value of $J$
in (\ref{Jarls}) for the strong CP angle $\theta \sim 0.8$.

We have now shown that even starting with a real \balpha,
a CP-violating phase in the CKM matrix of roughly the right order
of magnitude will be automatically generated.  What happens when
\balpha\ is allowed to be complex?  If \balpha\ has
complex elements but the phases of which do not change with scale,
then these phases will just cancel in taking inner products of
vectors, leading to exactly the same conclusions as for real
\balpha.  Indeed, this fact has been a major obstacle to 
our attempts at model building where we have never yet succeeded 
in generating an \balpha\ with a scale-dependent phase.  The
reason for this failure appears to be as follows.  Our models are 
based on the, to us, attractive assumption that fermion generation 
is dual to colour, which assumption guarantees that there are exactly 
3 generations of fermions as observed in experiment.  In that case, 
it seems natural that the rotation we want in generation space 
should have its origin in colour dynamics.  But if the QCD action 
is itself CP-conserving, then it cannot reasonably be expected to
give rotations in generation space producing in turn a mixing matrix 
which violates CP.  Although we have not yet proved that this is 
indeed the reason for our erstwhile failure, we have not succeeded 
either to obtain the contrary, and this is not without trying.  Given 
now, however, that the QCD action is actually CP-violating by virtue 
of the theta-angle term, it is natural to expect that CP-violating 
effects will result.  And this is what actually happens, although 
luckily not in the strong sector explicitly, so long as we take care 
to eliminate the theta-angle term from the strong action by the 
appropriate chiral transformations.  CP-violating effects will appear 
only in the weak sector where it is wanted.  Thus, the above result 
would seem also to have removed a major obstacle in our attempts at 
model building which, if substantiated, would be to us a great relief.

The conclusion at present is thus that with a rotating mass matrix
one seems on the one hand to be able to eliminate any theta-angle 
term by chiral transformations and yet retain a hermitian mass matrix 
and nonzero masses for all quarks, leaving the strong sector still
CP-invariant.  In the weak sector, on the other hand, the CKM mixing 
matrix appearing in the weak current acquires now an extra phase 
via the chiral transformations and leads to CP-violation there of 
roughly the order experimentally observed.  Thus, it seems to offer 
a solution simultaneously to both the strong and ``weak'' CP problems 
while, for the first time to our knowledge, linking the two together 
in a quite appealing manner, so that strong and weak CP-violations
appear now as just two different facets of one and the same phenomenon.  
Hence the theta-angle term in the QCD action in the strong sector need not 
be at all suppressed by nature, as is usually thought, but 
can instead
simply be re-expressed, via chiral transformations and mass matrix 
rotations, to reveal itself in the weak sector as the KM phase 
in the CKM mixing matrix, and to give rise to the familiar CP-violating 
effects observed there.  That this is the case adds much, we think, to 
the attractiveness of the rotation framework for explaining fermion 
mixing.  And as a bonus to us as model builders on the 
practical side, the new result has for the first time yielded a 
CP-violating phase in the CKM matrix which has so far eluded us in 
all the models that we have tried; this we find encouraging.


\begin{thebibliography}{99}

\bibitem{Weinberg} See for example: S. Weinberg, {\it The Quantum 
   Theory of Fields II} (Cambridge University Press, New York, 1996).

\bibitem{edm} C.A. Baker et al., {\it Phys. Rev. Lett.} {\bf 97}
   131801 (2006).

\bibitem{PecceiQuinn} R.D. Peccei and H. Quinn, {\it Phys. Rev. Lett.}
   {\bf 38}, 1440 (1977); {\it Phys. Rev. D} {\bf 16}, 1791 (1977).

\bibitem{Weinberg1} S. Weinberg, {\it Phys. Rev. Lett.} {\bf 40}, 223
   (1978).

\bibitem{Wilczek} F. Wilczek, {\it Phys. Rev. Lett.} {\bf 40}, 279 
   (1978).

\bibitem{KM} M. Kobayashi and K. Maskawa, {\it Prog. Theor. Phys.}
   {\bf 49}, 282 (1972).

\bibitem{Jarlskog} C. Jarlskog, {\it Z. Phys. C} {\bf 29} 491 (1985);
   {\it Phys. Rev. Lett.} {\bf 55}, 1039 (1985).

\bibitem{Databook} C. Amsler et al. (Particle Data Group), PL B667, 1 
    (2008) and 2009 partial update for the 2010 edition 
    (URL: http://pdg.lbl.gov).

\bibitem{strongcp} J. Bordes, Chan Hong-Mo and Tsou Sheung Tsun,
  arXiv:0707.3358, 
{\it Int. J. 
   Mod. Phys. A} {\bf 24}, 101 (2009). 

\bibitem{phenodsm} Chan Hong-Mo and Tsou Sheung Tsun, hep-th/9701120,
{\it Phys. Rev. D} 
{\bf 57}, 2507 (1998); 
J. Bordes, Chan Hong-Mo and Tsou Sheung Tsun, hep-ph/9901440,
{\it Eur. Phys.
   J. C} {\bf 10} 63 (1999).

\bibitem{Weinberg2} S. Weinberg, {\it Phys. Rev. D} {\bf 7}, 2887
   (1973).

\bibitem{Fritsch} H. Fritsch, {\it Nucl. Phys. B} {\bf 155}, 189
   (1978).

\bibitem{Harari} H. Harari, H. Haut, and J. Weyers, {\it Phys. Lett. B}
   {\bf 78}, 459 (1978)

\bibitem{cevidsm} J. Bordes, Chan Hong-Mo and Tsou Sheung Tsun,
  hep-ph/0203124,  
{\it Eur. Phys. J. C} {\bf 27}, 189 (2003).

\bibitem{genmixdsm} Chan Hong-Mo and Tsou Sheung Tsun, hep-ph/0303010,
{\it Acta Phys. Pol.}
   {\bf 33}, 4041 (2002)

\bibitem{prepsm} Chan Hong-Mo and Tsou Sheung Tsun, arXiv:hep-ph/0611364,
{\it Eur. Phys. J. C} {\bf 52} 635 (2007).

\bibitem{features} J. Bordes, Chan Hong-Mo, 
J. Pfaudler, and Tsou Sheung Tsun,  hep-ph/9802436,
   {\it Phys. Rev. D} {\bf 58}, 053006 (1998).


\end{thebibliography}
\end{document}